\documentclass[english]{article}
\usepackage{amsmath}
\usepackage{amssymb}
\usepackage{cite}

\def\equ#1{(\ref{#1})}
\usepackage{epsfig}
\makeatletter

\usepackage{babel}

\begin{document}
\begin{titlepage}
\thispagestyle{empty}

\bigskip

\begin{center}
\noindent{\Large \textbf
{Bosonic Fields in Crystal Manyfold}}\\

\vspace{0,5cm}

\noindent{G. Alencar ${}^{a}$\footnote{e-mail: geovamaciel@gmail.com }, R. R. Landim ${}^{a}$\footnote{e-mail: renan@fisica.ufc.br}, M. O. Tahim ${}^{b}$\footnote{e-mail: makarius.tahim@gmail.com } and R.N. Costa Filho ${}^{a}$\footnote{e-mail: rai@fisica.ufc.br }}

\vspace{0,5cm}
 
 {\it ${}^a$Departamento de F\'{\i}sica, Universidade Federal do Cear\'{a}-
Caixa Postal 6030, Campus do Pici, 60455-760, Fortaleza, Cear\'{a}, Brazil. 
 }
 
{\it ${}^b$Universidade Estadual do Cear\'a, Faculdade de Educa\c c\~ao, Ci\^encias e Letras do Sert\~ao Central- 
R. Epitcio Pessoa, 2554, 63.900-000  Quixad\'{a}, Cear\'{a},  Brazil.
 
\vspace{0.2cm}
 }
 
\end{center}

\vspace{0.3cm}

\begin{abstract}
A chain-like or a 1D crystal-like universe made of intersecting membranes in extra dimensions in a Randall-Sundrum scenario is studied. A background gravitational metric satisfying the right boundary conditions is considered to study the 
localization of the scalar, gauge and Kalb-Ramond field. It is found that the wave function for the fields are Bloch waves. The mass modes equations are calculated allowing us to show the zero-gap mass behavior and the mass dispersion relation 
for each field. Finally we generalize all these results and consider $q-$forms in the crystal membrane universe. We show that, depending to the dimension $D$, the form $q$ and the dilaton coupling $\lambda$ the mass spectrum can be the same for 
the different bosonic field studied. 
\end{abstract}

\end{titlepage}

\section{Introduction}

When studying aspects related to localization of fields in membranes, we necessarily
need to understand the solutions of differential equations with eingenvalues. This
eventually leads to the study a Schr\"odinger like equation openning  the possibility for interpretation and comparison between methods and aspects of physics
of extra dimensions and those of condensed matter physics. In some sense, it is another way
to link these two areas, despite the celebrated AdS/CDM duality. Of course, there is no duality
linking two theories with different regimens of coupling constant, but the basic mathematics of particles subject to potential barriers is just the same as those we attack in our
training through the several famous quantum mechanical books. The question is how far can we go
in this comparison. Besides the most basic example, the process of localization of fields (the zero modes or bound states), there is the study of resonances where the massive spectrum of particles interact with the membranes through the potential they 
generate \cite{Bazeia:2007nd,Fonseca:2012bw,Chumbes:2011zt,Cunha:2011yk,Almeida:2009jc,Liu:2009ve,Liu:2009mga,Cruz:2009ne,Zhao:2010mk,Li:2010dy,Cruz:2011kj,Alencar:2012en,Landim:2011ts,Landim:2011ki}. This kind of model provides a possible 
solution to the hierarchy problem and show how gravity is trapped to a membrane and also has
important phenomenological implications to the standard model and cosmology \cite{Csaki:2002gy,Csaki:1999mp, Csaki:1999jh}. Another very interesting step in that direction is the model regarding a universe with extra dimensions in such a way that the 
membranes itself generate a kind of chain or 1D-crystal  \cite{Kaloper:1999et,Kaloper:2004cy,Oda:1999di}. In such situation we can think about membranes in every direction
in the bulk, with intersections. These are configurations already discussed in the context of string theory phenomenology, where we can generate families of particles of the standard model plus some 
corrections \cite{Aldazabal:2000cn,Ibanez:2001nd,Blumenhagen:2001te,Blumenhagen:2005mu,Lust:2004ks}. Moreover, in string theory  and supergravity the antisymmetric tensor fields arise 
naturally \cite{Polchinski:1998rq,Polchinski:1998rr,VanNieuwenhuizen:1981ae} and play an important role in dualization \cite{Smailagic:1999qw,Smailagic:2000hr}. In particular, they appear in the $R-R$ sector 
of each of the type II string theories. These tensor fields couple naturally to higher-dimensional extended objects, the so-called $D-$branes, and are important to make them stable. Besides this, they are 
related to the linking number of higher dimensional knots \cite{Oda:1989tq} and to the mechanism of topological mass generation \cite{Landim:2000ir,Landim:2001zu}. The rank of these antisymmetric tensors is defined by the dimension of the manifold \cite{Nakahara:2003nw}. Beyond that, these kind of fields also 
play an important role in the solution of the moduli stabilization problem of string theory \cite{Kachru:2002he,Antoniadis:2004pp,Balasubramanian:2005zx}. Because of these aspects, it is important to study 
higher rank tensor fields in membrane backgrounds. In this direction antisymmetric tensor fields have already been considered in models of extra dimension. 

In general the $q-$forms of highest rank do not have physical relevance because when the rank increases the number of gauge freedom also increases. However, that can be used to cancel the dynamics of the field in the brane \cite{Alencar:2010vk}, 
leading to the study of the mass spectrum of the two and three-form, for example, in refs. \cite{Mukhopadhyaya:2004cc} and \cite{Mukhopadhyaya:2007jn} in a context of five dimensions with codimension one. The coupling between the two and 
three-forms with the dilaton was studied, in different contexts, in \cite{DeRisi:2007dn,Mukhopadhyaya:2009gp,Alencar:2010mi}. In these scenarios, some facts about localization of fields are known: the scalar field ($0-$form) is localizable, 
but the vector gauge fields ($1-$form), the Kalb-Ramond field($2-$form) and the three-form field are not. That happens because in four dimensions the vector field is conformal and all information coming from warp factors drops out, 
necessarily rendering a non-normalizable four dimensional 
effective action. Beyond that,  Kehagias and Tamvakis \cite{Kehagias:2000au} showed that the coupling between the dilaton and the vector gauge field produces localization of the vector field, and Alencar {\it et.al.,} considered these 
coupling with a three-form field where a condition for localization was found \cite{Alencar:2010hs}. More recently, the problem has been generalized to $q$-forms \cite{Landim:2010pq,Fu:2012sa}.

The above mentioned facts have influence on the phenomenology of the four dimensional world we live \cite{ArkaniHamed:1999zg}. The point we would like to address here is that, in the case of membranes distributed along extra dimensions which are 
slices of AdS spaces, we can find Bloch like solutions to the equations of motion. Therefore, one can argue that these models can present a similar behavior of quasi-particles like phonons, magnons, or electrons in the presence of a periodic potential. 
The goal here is to extend previous works in this subject to study the behavior of bosonic form fields in the crystal manyfold universe. We go right to the point of analysis of the fields subject to the background described in \cite{Kaloper:1999et}. 
With that in mind, the organization of the paper is as follows. In the first section we shortly review the background of the crystal manyfold universe. In the second section we make a preliminary discussion of the general Schr\"odinger like equations 
we must 
treat. In the third, fourth, fiftieth and sixtieth sections we discuss the behavior of the scalar field, the gauge vector field, the Kalb-Ramond field and general $q$-forms, respectively. We discuss for each case the Bloch like behavior of the wave 
functions and give numerical results for mass bands. In the last section we present conclusions, more discussions and perspectives for future works.

\section{The Crystal Manyfold Background}

In this section we give a short review of the background metric obtained
in the crystal manyfold. We only consider the simplest case with one extra
dimension. Following ref. \cite{Kaloper:1999et}, the conformal metric is defined as
\begin{equation}\label{metricaconforme}
ds_{5}^{2}=\Omega^{2}(z)(\eta_{\mu\nu}dx^{\mu}dx^{\nu}+dz{}^{2}),
\end{equation}
with
\begin{equation}
\Omega^{-1}=K{\cal S}(z)+1
\end{equation}
where in the above equation $K=(L)^{-1}$ ($L$ is the AdS radius). The function $\ensuremath{{\cal S}(z)}$
satisfy the equations
\begin{eqnarray*}
 &  & \frac{d^{2}{\cal S}(z)}{d(z)^{2}}=2\sum_{j}(-1)^{j}\delta(z-jl),\\
&&\left |\frac{d{\cal S}(z)}{dz}\right |=1,
\end{eqnarray*}
and are given by ($l$ is the spacing between the branes)
\begin{equation}
{\cal S}(z) = \begin{cases} ...&\\
2pl - z, & \text{for $(2p-1) l < z< 2p l$;}\\
z-2pl, &  \text{for $2pl< z < (2p+1) l$;}\\
...\quad .&
\end{cases}
\end{equation}
In order to analyze the dilaton with a conformal metric \equ{metricaconforme},
we will use the same procedure as in ref. \cite{Alencar:2012en} where the metric
with dilaton is 
\begin{equation}
\label{metricadilaton}
ds^{2}=e^{2A(y)}\eta_{\mu\nu}dx^{\mu}dx^{\nu}+e^{2B(y)}dy^{2},
\end{equation}
and after solving Einstein's equation we get the relations \cite{Kehagias:2000au}.
\begin{equation}
B(y)=A(y)/4; \qquad \pi=-\sqrt{3M^{3}}A(y).
\end{equation}
We introduce the parameter $b$ such that $B(y)=(1-b)A(y)$, where $b=3/4$ is the
case with dilaton and $b=1$ otherwise.
To obtain the conformal metric \equ{metricaconforme}, we use the transformation
$dy/dz=e^{A(y)-B(y)}=e^{bA(y)}=\Omega(z)$ in \equ{metricadilaton}. We obtain
$\bar{A}(z)=\ln\Omega(z)/b$, where $A(y)=\bar{A}(z)$.

\section{The General Case}

Here we focus on the prototype Schr\"odinger-like equation that will appear through all this work. We
consider the solution with some details since it will be important
for all cases considered. As mentioned before we can relate the
potentials with the ones obtained previously in \cite{Alencar:2012en}. In all
the cases the equation of motion has the following general form
\begin{equation}
\left[-\frac{d^{2}}{dy^{2}}+P'(y)\frac{d}{dy}+V(y)\right]\psi(y)=m^{2}Q(y)\psi(y),
\end{equation}
where $P(y)=\gamma A(y)$, $Q(y)=e^{-2bA(y)}$ and $V(y)=0$ for all
fields except gravity, in which $V(y)=2A''(y)-2(1+b)A'(y)^{2}$. We
can transform this into a Schr\"odinger-like equation through the transformations
\begin{equation}
\frac{dz}{dy}=f(y),\quad\psi(y)=\Theta(y)\overline{\psi}(z),\label{transformation}
\end{equation}
with
\begin{equation}
f(y)=\sqrt{Q(y)},\quad\Theta(y)=\exp(P(y)/2)Q(y)^{-1/4}.
\end{equation}
As stressed before, when we have an analytical expression for $dy/dz$
the potential is given by
\begin{equation}
\bar{U}(z)=\bar{V}(z)/\bar{f}^{2}(z)+\frac{\bar{P}'(z)\bar{\Theta}'(z)-\bar{
\Theta}''(z)}{\bar{\Theta}(z)}
-\frac{\bar{\Theta}'(z)}{\bar{\Theta}(z)}\frac{\bar{f}'(z)}{\bar{f}(z)}
\end{equation}
where $f(y)=\bar{f}(z)$. With this we arrive to a Schr\"odinger like equation
\begin{equation}
 \left[-\frac{d^2}{dz^{2}}+\bar{U}\right]\bar{\psi}(z)=m^{2}\bar{\psi}(z),
\end{equation}
where the potential  $\bar{U}(z)$ is given by
\begin{equation}
 \bar{U}(z)=c\bar{A}^{''}(z)+c^{2}[\bar{A}'(z)]^{2}.
\end{equation}
Now, considering $\bar{A}(z) =\ln\Omega(z)/b$. We
get 
\begin{equation}
\bar{U}(z)=\left(\frac{c}{b}+\frac{c^{2}}{b^{2}}\right)\frac{(\Omega^{-1})'^{2}}{(\Omega^{
-1 } )^ { 2}}-\frac{c}{b}\frac{(\Omega^{-1})''}{(\Omega^{-1})}.
\end{equation}

The potential has been expressed in terms of $\Omega^{-1}$ due to
the form of the conformal factor. It is interesting  to note that the form of the potential coefficients simplifies the solution. Taking $\Omega$
given before we get the final Schr\"odinger-like equation
\begin{equation}
\psi''+\left[m^{2}-\left(\frac{c}{b}+\frac{c^{2}}{b^{2}}\right)\frac{1}{({\cal S}(z)+L)^{2}}+2\frac{c}{b}\frac{\sum_{j}(-1)^{j}\delta(z-jl)}{({\cal S}(z)+L)}\right]\psi=0.
\end{equation}
This equation is very similar to the Schr\"odinger equation for a particle 
propagating in a one-dimensional periodic potential (1D-crystal). In order to solve it we have to consider two adjacent elementary cells in the region $0\leq z\leq4l$. Focusing
on the region $0\leq z\leq2l$ we need to find the appropriate boundary
conditions. First of all we must have $\psi(l_{+})=\psi(l_{-})$ and $\psi(2l_{+})=\psi(2l_{-})$. Where we use the definition $l_{\pm}=\lim_{\epsilon\to0}(l\pm\epsilon)$.
To obtain the correct boundary condition for the derivative we simply
integrate the above equation from $l_{-}(2l_{-})$ to $l_{+}(2l_{+})$ to
obtain
\begin{eqnarray}
\psi'(l_{+})-\psi'(l_{-})&=&\frac{2c}{b(l+L)}\psi(l);\nonumber \\
\psi'(2l_{+})-\psi'(2l_{-})&=&-\frac{2c}{bL}\psi(2l).
\end{eqnarray}
The equation for the first elementary cell is
\begin{equation}
\psi''+m^{2}\psi  = \begin{cases}
\frac{(\frac{c}{b}+\frac{c^{2}}{b^{2}})}{(z+L)^{2}}\psi\quad for \qquad  0<z<l\\
\frac{(\frac{c}{b}+\frac{c^{2}}{b^{2}})}{(2l-z+L)^{2}}\psi\quad for \qquad  l<z<2l
\end{cases}
\end{equation}
and the above equations can be expressed in terms of a Bessel equation.
For this we just perform the transformations
\begin{eqnarray}
\psi&=&
\begin{cases}
\sqrt{u}\Psi(u), & u=m(z+L),\quad for\qquad 0<z<l;\\
\sqrt{v}\Psi(v), & v=m(2l-z+L),\quad for \qquad l<z<2l,
\end{cases}
\end{eqnarray}
to get
\begin{equation}
u^{2}\Psi''+u\Psi'+(u^{2}-\nu^{2})\Psi=0,
\end{equation}
where $\nu^{2}=(\frac{1}{2}+\frac{c}{b})^{2}$ and the same equation
is valid for the variables $u$ and $v$. It is important to mention that the quantity $\nu^{2}$ is related to the kind of field we consider as we will show later.  Therefore the solution of the above
equation is given by
\begin{eqnarray}
\psi &=&
\begin{cases}
\sqrt{u}(AH_{\nu}^{+}(u)+BH_{\nu}^{-}(u)), & 0<z<l,\qquad u=m(z+L);\\
 \sqrt{v}(CH_{\nu}^{+}(v)+DH_{\nu}^{-}(v)),&l<z<2l,\qquad v=m(2l-z+L).
\end{cases}
\end{eqnarray}

Now using the transfer matrix technique we can relate the constants $C,D$ with $A,B$. In order to impose boundary conditions
for the wave function and for its first derivative we define
\begin{eqnarray}
E&=&\sqrt{u}H_{\nu}^{+}(u),E(l)=\sqrt{m(l+L)}h_{\nu}^{+}; \nonumber
\\
F&=&\sqrt{u}H_{\nu}^{-}(u),F(l)=\sqrt{m(l+L)}h_{\nu}^{-},
\end{eqnarray}
and
\begin{eqnarray}
G&=&\sqrt{v}H_{\nu}^{+}(v),\qquad G(l)=E(l);\nonumber
\\
I&=&\sqrt{v}H_{\nu}^{-}(v),\qquad I(l)=F(l),
\end{eqnarray}
with the following properties
\begin{eqnarray}
\frac{d}{dz}E(u)|_{z=l}&=&mE'(l)=-mG'(l);\\ \nonumber
\frac{d}{dz}F(u)|_{z=l}&=&mF'(l)=-mI'(l),
\end{eqnarray}
where the prime means a derivative with respect to the argument. After a bit of algebra
we obtain
\begin{eqnarray}
\begin{pmatrix}\psi(u)\\
\frac{d}{dz}\psi(u)
\end{pmatrix}&=&\begin{pmatrix}E & F\\
mE' & mF'
\end{pmatrix}\begin{pmatrix}A\\
B
\end{pmatrix},\nonumber
\\
\begin{pmatrix}\psi(v)\\
\frac{d}{dz}\psi(v)
\end{pmatrix}&=&\begin{pmatrix}G & I\\
-mG' & -mI'
\end{pmatrix}\begin{pmatrix}C\\
D
\end{pmatrix}.
\end{eqnarray}
With this the boundary conditions can be accounted by imposing

\begin{eqnarray}
&&
\begin{pmatrix}E(l) & F(l)\\
mE'(l)+\frac{2c}{b(l+L)}E(l) & mF'(l)+\frac{2c}{b(l+L)}F(l)
\end{pmatrix}\begin{pmatrix}A\\
B
\end{pmatrix}\nonumber
\\
&&=\begin{pmatrix}E(l) & F(l)\\
-mE'(l) & -mF'(l)
\end{pmatrix}\begin{pmatrix}C\\
D
\end{pmatrix}.
\end{eqnarray}
Inverting the matrix in the RHS
\begin{equation}
 \begin{pmatrix}E(l) & F(l)\\
-mE'(l) & -mF'(l)
\end{pmatrix}^{-1}=\frac{1}{m^{2}(l+L)(h_{\nu}^{-}h_{\nu-1}^{+}-h_{\nu}^{+}h_{\nu-1}^{-})}\begin{pmatrix}-mF'(l) & -F(l)\\
+mE'(l) & E(l)
\end{pmatrix},
\end{equation}
we obtain
\begin{equation}
\begin{pmatrix}C\\
D
\end{pmatrix}=\begin{pmatrix}-\frac{h_{\nu}^{-}h_{\nu-1}^{+}+h_{\nu}^{+}h_{\nu-1}^{-}}{h_{\nu}^{-}h_{\nu-1}^{+}-h_{\nu}^{+}h_{\nu-1}^{-}} &
\frac{-2h_{\nu}^{-}h_{\nu-1}^{-}}{h_{\nu}^{-}h_{\nu-1}^{+}-h_{\nu}^{+}h_{\nu-1}^{-}} \\
\frac{2h_{\nu}^{+}h_{\nu-1}^{+} }{h_{\nu}^{-}h_{\nu-1}^{+}-h_{\nu}^{+}h_{\nu-1}^{-}} &
\frac{(h_{\nu}^{-}h_{\nu-1}^{+}+h_{\nu}^{+}h_{\nu-1}^{-})}{h_{\nu}^{-}h_{\nu-1}^{+}-h_{\nu}^{+}h_{\nu-1}^{-}}
\end{pmatrix}\begin{pmatrix}A\\
B
\end{pmatrix}\equiv K\begin{pmatrix}A\\
B
\end{pmatrix}.
\end{equation}
Similarly, we can solve for the next cell ($2l<z<4l$) to find
\begin{equation}
\begin{pmatrix}\hat{C}\\
\hat{D}
\end{pmatrix}=\begin{pmatrix}-\frac{h_{\nu}^{-}h_{\nu-1}^{+}+h_{\nu}^{+}h_{\nu-1}^{-}}{h_{\nu}^{-}h_{\nu-1}^{+}-h_{\nu}^{+}h_{\nu-1}^{-}} &
\frac{-2h_{\nu}^{-}h_{\nu-1}^{-}}{h_{\nu}^{-}h_{\nu-1}^{+}-h_{\nu}^{+}h_{\nu-1}^{-}} \\
\frac{2h_{\nu}^{+}h_{\nu-1}^{+} }{h_{\nu}^{-}h_{\nu-1}^{+}-h_{\nu}^{+}h_{\nu-1}^{-}} &
\frac{(h_{\nu}^{-}h_{\nu-1}^{+}+h_{\nu}^{+}h_{\nu-1}^{-})}{h_{\nu}^{-}h_{\nu-1}^{+}-h_{\nu}^{+}h_{\nu-1}^{-}}
\end{pmatrix}\begin{pmatrix}\hat{A}\\
\hat{B}
\end{pmatrix}= K\begin{pmatrix}\hat{A}\\
\hat{B}
\end{pmatrix}
\end{equation}
and therefore we have four constants $A,B,\hat{A}$and $\hat{B}$
to be determined. In order to get this we impose the following conditions
\begin{eqnarray}
\psi(2l) & = & e^{2iql}\psi(0),\\
\psi(4l) & = & e^{2iql}\psi(2l),\\
\psi(2l_{+}) & = & \psi(2l_{-}),\\
\psi'(2l_{+})-\psi'(2l_{-}) & = & -\frac{2c}{bL}\psi(2l).
\end{eqnarray}
The first of the above conditions can be written as
\begin{equation}
e^{2iql}\begin{pmatrix}\hat{h}_{\nu}^{+} & \hat{h}_{\nu}^{-}\end{pmatrix}\begin{pmatrix}A\\
B
\end{pmatrix}=\begin{pmatrix}\hat{h}_{\nu}^{+} & \hat{h}_{\nu}^{-}\end{pmatrix}\begin{pmatrix}C\\
D
\end{pmatrix}=\begin{pmatrix}\hat{h}_{\nu}^{+} & \hat{h}_{\nu}^{-}\end{pmatrix}K\begin{pmatrix}A\\
B
\end{pmatrix},
\end{equation}
where we have used $H(2l)=H(0)\equiv\hat{h}$, resulting in 
\begin{equation}\label{B}
B=-\frac{\hat{h}_{\nu}^{+}K_{11}+\hat{h}_{\nu}^{-}K_{21}-e^{2iql}\hat{h}_{\nu}^{+}}{\hat{h}_{\nu}^{+}K_{12}+\hat{h}_{\nu}^{-}K_{22}-e^{2iql}\hat{h}_{\nu}^{-}}A\equiv fA
\end{equation}
and identically for the second condition
\begin{equation}\label{hatB}
\hat{B}=-\frac{\hat{h}_{\nu}^{+}K_{11}+\hat{h}_{\nu}^{-}K_{21}-e^{2iql}\hat{h}_{\nu}^{+}}{\hat{h}_{\nu}^{+}K_{12}+\hat{h}_{\nu}^{-}K_{22}-e^{2iql}\hat{h}_{\nu}^{-}}\hat{A}= f\hat{A}.
\end{equation}
From the third condition we get
\begin{equation}
\hat{A}\hat{h}_{\nu}^{+}+\hat{B}\hat{h}_{\nu}^{-}=e^{2iql}(A\hat{h}_{\nu}^{+}+B\hat{h}_{\nu}^{-})
\end{equation}
and using the last two relations we find
\begin{equation}\label{hatA}
\hat{A}=e^{2iql}A.
\end{equation}
The last two conditions are identical to the contour conditions used
in $z=l$ and $z=3l$, therefore from them we obtain
\begin{equation}
\begin{pmatrix}\hat{A}\\
\hat{B}
\end{pmatrix}=\hat{K}\begin{pmatrix}C\\
D
\end{pmatrix}=\hat{K}K\begin{pmatrix}A\\
B
\end{pmatrix}=\hat{K}K\begin{pmatrix}A\\
fA
\end{pmatrix}
\end{equation}
where $\hat{K}$ is the same matrix as $K$ with $h\to\hat{h}$. Using
now the Eqs. (\ref{B}),(\ref{hatB}) and (\ref{hatA}) we get
\begin{equation}
\begin{pmatrix}e^{2iql}\\
fe^{2iql}
\end{pmatrix}=\hat{K}K\begin{pmatrix}1\\
f
\end{pmatrix}
\end{equation}
and finally we arrive to 
\begin{eqnarray}\label{massband}
\cos(lq)&=&\frac{(j_{\nu}n_{\nu-1}+j_{\nu-1}n_{\nu})(\hat{j}_{\nu}\hat{n}_{\nu-1}+\hat{j}_{\nu-1}\hat{n}_{\nu})-\hat{j}_{\nu-1}\hat{j}_{\nu}(j_{\nu-1}j_{\nu}+3n_{\nu-1}n_{\nu})}{2(j_{\nu}n_{\nu-1}-j_{\nu-1}n_{\nu})(\hat{j}_{\nu}\hat{n}_{\nu-1}-\hat{j}_{\nu-1}\hat{n}_{\nu})} \nonumber \\
&-&\frac{\hat{n}_{\nu-1}\hat{n}_{\nu}(3j_{\nu-1}j_{\nu}+n_{\nu-1}n_{\nu})}{2(j_{\nu}n_{\nu-1}-j_{\nu-1}n_{\nu})(\hat{j}_{\nu}\hat{n}_{\nu-1}-\hat{j}_{\nu-1}\hat{n}_{\nu})}.
\end{eqnarray}
The above equation gives us a dispersion relation  for the mass, where one can
find, for example, the mass gap of the system considered. In order to obtain numerical results
we have to discuss the magnitude of the parameters involved. We have that $mc^{2}= (\hbar c)\frac{1}{l}$
where we can compare units of mass and length. For this purpose we use $\hbar c\simeq 200 MeV\times 10^{-15}m$. In the case where $l>L$, the corrections for the newtonian potential for particles localized to a $3$-brane at the intersections of branes with negative tension is approximately given by \cite{Kaloper:1999et}
 
\begin{equation}
V_{N}\simeq - G_{N}\frac{M_{1}M_{2}}{r}\left(1+\frac{al^{3}}{L^2}\frac{e^{-br/l}}{r}\right), 
\end{equation}
where $a$ and $b$ are constants of order unity. With that we can
have gravity in the observational bounds, i.e., $\frac{l^{3}}{L^2}\leq 1mm$. For this case 
$M^{2}_{PL}\sim M^{3}_{*}L\aleph$ ($\aleph$ is the number of branes in the crystal and $L$ is the AdS radius).
The fundamental scale should satisfy $M_{*}\sim\frac{l}{L}\cdot TeV$. If
$l\sim eV^{-1}\ll 1mm$ we find $\aleph\sim 10^{16}$ branes and $M_{*}\sim 100 TeV$ ($\frac{l}{L}\sim 100$).
That is the result we put in our numerical analysis. In the next sections we present the results for each case
of bosonic fields in detail. 

\section{Scalar Field on the Crystal Manyfold}

In this section we consider the scalar field case. Before analyzing
the massive modes we must look for the localization of the zero mode.
The action for the scalar field without the dilaton coupling is given
by
\begin{equation}
S=\int d^{4}xdz\sqrt{-G}G^{MN}\partial_{M}\Phi\partial_{N}\Phi
\end{equation}
and the zero mode is obtained by considering $\Phi=\chi(x)$, which
is equivalent to set $m=0$. Considering the metric in the form of Eq. (\ref{metricaconforme}) we get for the effective action
\begin{equation}
S=\int dz\Omega^{3}(z)\int d^{4}x\eta^{\mu\nu}\partial_{\mu}\chi\partial_{\nu}\chi,
\end{equation}
and according to the solution $\int\Omega^{3}(z)<\infty$ we have
a well defined four dimensional action. It is clear that the scalar field
is localized without the necessity of the dilaton coupling. This coupling
will not change the localizability of the field, but it must be considered because of the
gauge field localization. Therefore, due to consistency it will be include in the analysis. 

To find the massive modes we should solve the
equation of motion. Here we use the metric defined by Eq. (\ref{metricadilaton}). As we have found before, the potential is written in a general form
 \begin{equation}
 \bar{U}(z)=c_i\bar{A}^{''}(z)+c_i^{2}[\bar{A}'(z)]^{2}.
\end{equation} 
The prime means a derivative with respect to $z$, and to obtain
that result we have used $dy/dz=e^{-b_iA}$, where the label $i=1$ and $2$ stands for the case with and without the dilaton, respectively, and are given by
\begin{eqnarray}
b_{1}&=&\frac{3}{4},\quad c_{1}=\left(\frac{3}{2}+\frac{\lambda\sqrt{3M^{3}}}{2}\right);\nonumber
\\
b_{2}&=&1,\quad c_{2}=\frac{3}{2},
\end{eqnarray}
and we get
\begin{eqnarray}
\nu_{1}^{2}&=&\left(\frac{5}{2}+\frac{2\lambda\sqrt{3M^{3}}}{3}\right)^{2};\nonumber
\\
\nu_{2}&=&2.
\end{eqnarray}

\section{The Gauge Field in Crystal Manyfold}

Now we consider the gauge field living in the crystal manyfold. The action
for this field is given by
\begin{equation}
S_{X}=\int d^{5}x\sqrt{-G}[Y_{M_{1}M_{2}}Y^{M_{1}M_{2}}],
\end{equation}
where $Y_{M_{1}M_{2}}=\partial_{[M_{1}}X_{M_{2}]}$ is the field strength
for the $1-$form $X$. The study of the zero mode is made using
$X_{M_{1}}=X_{M_{1}}(x^{\mu})$ and, as said in the second section, we must
consider the metric given in Eq. (\ref{metricaconforme})
for the case without the dilaton coupling. In this case we get the effective
action
\begin{equation}
S_{X}=\int dz\Omega\int d^{4}x[Y_{\mu_{1}\mu_{2}}Y^{\mu_{1}\mu_{2}}].
\end{equation}
Like in the case with just one brane $\int dz\Omega=\infty$,
and we have a ill defined action. However, we can consider the coupling
of this field with the dilaton, with action given by
\begin{equation}
S_{X}=\int d^{5}x\sqrt{-G}e^{-\lambda\pi}[Y_{M_{1}M_{2}}Y^{M_{1}M_{2}}],
\end{equation}
where $\lambda$ is the dilaton coupling. With this and using the metric defined in Eq. (\ref{metricadilaton}) the effective
action becomes
\begin{equation}
S_{X}=\int dz\Omega^{(\frac{5}{4}-\lambda\pi)}\int d^{4}x[Y_{\mu_{1}\mu_{2}}Y^{\mu_{1}\mu_{2}}].
\end{equation}
The integration must be performed in the crystal manyfold. However,
as mentioned previously, the finiteness of the $z$ dependence is reduced to
the finiteness of integral in one cell. With the expression for $\Omega$
we see that this is reached if $\lambda>-1/\sqrt{3M^{3}}$. This is the same condition as that obtained for the localization of gauge field in the model with just one brane. In order to analyze the massive
modes, we must come back to the metric defined by Eq. (\ref{metricadilaton}).
The equations of motion are given by
\begin{equation}
\partial_{M_{1}}(\sqrt{-G}G^{M_{1}P}G^{M_{2}Q}Y_{PQ})=0.
\end{equation}

We use a gauge freedom to fix $X_{y}=\partial^{\mu}X_{\mu}=0$ and by
using the transformations $\frac{dz}{dy}=e^{-3A/4}$ and $\frac{dz}{dy}=e^{-A}$
respectively, we reach the potentials for the Schr\"odinger-like equation
for the cases with and without the dilaton as before

 \begin{equation}
 \bar{U}(z)=c_i\bar{A}^{''}(z)+c_i^{2}[\bar{A}'(z)]^{2},
\end{equation} 
where now
\begin{eqnarray}
b_{1}&=&\frac{3}{4},\quad c_{1}=\left(\frac{1}{2}+\frac{\lambda\sqrt{3M^{3}}}{2}\right),\nonumber
\\
b_{2}&=&1,\quad c_{2}=\frac{1}{2},
\end{eqnarray}
and we get  
\begin{eqnarray}
\nu_{1}^{2}&=&\left(\frac{7}{6}+\frac{2\lambda\sqrt{3M^{3}}}{3}\right)^{2},\nonumber
\\
\nu_{2}&=&1.
\end{eqnarray}

\section{The Kalb-Ramond Field in Crystal Manyfold}

Now we turn our attention to the Kalb-Ramond field. The action for
this field is defined by
\begin{equation}
S_{X}=\int d^{5}x\sqrt{-G}Y_{M_{1}M_{2}M_{3}}Y^{M_{1}M_{2}M_{3}},
\end{equation}
where $Y_{M_{1}M_{2}M_{3}}=\partial_{[M_{1}}X_{M_{2}M_{3}]}$ is the
field strength for the $2-$form $X$. Again, we must consider the
zero mode given by $X_{M_{1}M_{2}}=X_{M_{1}M_{2}}(x^{\mu})$, with an effective action in the conformal metric (\ref{metricaconforme}) given by
\begin{equation}
S_{X}=\int dz\Omega^{-2}\int d^{4}x[Y_{\mu_{1}\mu_{2}\mu_{3}}Y^{\mu_{1}\mu_{2}\mu_{3}}],
\end{equation}
and, as in the gauge field case, we have a ill defined effective action
since $\int dz\Omega^{-2}=\infty$. Now we turn to the action with the dilaton coupling defined by
\begin{equation}
S_{X}=\int d^{5}x\sqrt{-G}e^{-\lambda\pi}Y_{M_{1}M_{2}M_{3}}Y^{M_{1}M_{2}M_{3}},
\end{equation}
where again $\lambda$ is the dilaton coupling. With this and using the metric in Eq. (\ref{metricadilaton}) the effective action becomes
\begin{equation}
S_{X}=\int dz\Omega^{(-\frac{3}{4}-\lambda\pi)}\int d^{4}x[Y_{\mu_{1}\mu_{2}}Y^{\mu_{1}\mu_{2}}].
\end{equation}
With the expression for $\Omega$ we see that the zero mode localization is reached if $\lambda>7/4\sqrt{3M^{3}}$. Again we find the same condition as that obtained for the localization
of Kalb-Ramond field in the model with just one brane. Now we analyze
the massive modes coming back to the metric defined by Eq. (\ref{metricadilaton}).
The new equation of motion is:
\begin{equation}
\partial_{M}(\sqrt{-G}G^{MP}G^{NQ}G^{LR}e^{-\lambda\pi}H_{PQR})=0.
\end{equation}
Here, we can use gauge freedom to fix $X_{\mu_{1}y}=\partial^{\mu_{1}}X_{\mu_{1}\mu_{2}}=0$
and by using the transformations $\frac{dz}{dy}=e^{-3A/4}$ and $\frac{dz}{dy}=e^{-A}$
respectively, we have the following expression for the constants $b_i$ and $c_i$
\begin{eqnarray}
b_{1}&=&\frac{3}{4},\quad c_{1}=\left(-\frac{1}{2}+\frac{\lambda\sqrt{3M^{3}}}{2}\right),\nonumber
\\
b_{2}&=&1,\quad c_{2}=-\frac{1}{2},
\end{eqnarray}
and we obtain the solutions
\begin{eqnarray}
\nu_{1}^{2}&=&\left(-\frac{1}{6}+\frac{2\lambda\sqrt{3M^{3}}}{3}\right)^{2},\nonumber
\\
\nu_{2}&=&0.
\end{eqnarray}

\section{The $q-$Form Field in Crystal Manyfold}

Now we can readily generalize our previous results to a $q-$form in
a $p-$brane, where $p=D-2$. In a recent paper the authors have
considered this issue and we are going to use those results here \cite{Landim:2010pq}. Just
as before we must consider the cases with and without the dilaton
coupling. The action is given by
\begin{equation}
S_{X}=\int d^{D}x\sqrt{-G}[Y_{M_{1}...M_{q+1}}Y^{M_{1}...M_{q+1}}]
\end{equation}
and considering $X_{M_{1}M_{2}M_{3}}=X_{M_{1}M_{2}M_{3}}(x^{\mu})$
with the metric of Eq. (\ref{metricaconforme}) we arrive at the effective action
\begin{equation}
S_{X}=\int dz\Omega^{D-2(q+1)}\int d^{4}x[Y_{\mu_{1}...\mu_{q+1}}Y^{\mu_{1}...\mu_{q+1}}].
\end{equation}

It is evident why the gauge field is not localizable in five dimensions.
From the above we get the condition $q<(D-3)/2$, which for $D=5$
give us $q<1$. We also see that in higher dimensions we can have
localized form fields without the inclusion of the dilaton coupling.
However, if we want to localize all the form fields we must consider
the dilaton coupling with an action defined by
\begin{equation}
S_{X}=\int d^{D}x\sqrt{-G}e^{-\lambda\pi}[Y_{M_{1}...M_{q+1}}Y^{M_{1}...M_{q+1}}]
\end{equation}
and considering again $X_{M_{1}M_{2}M_{3}}=X_{M_{1}M_{2}M_{3}}(x^{\mu})$ and Eq. (\ref{metricadilaton})
we get the effective action
\begin{equation}
S_{X}=\int dz\Omega^{(p-2q+\frac{1}{4}-\lambda\pi)}\int d^{4}x[Y_{\mu_{1}...\mu_{q+1}}Y^{\mu_{1}...\mu_{q+1}}].
\end{equation}
\begin{figure}[t]
\centerline{\epsfig{file= 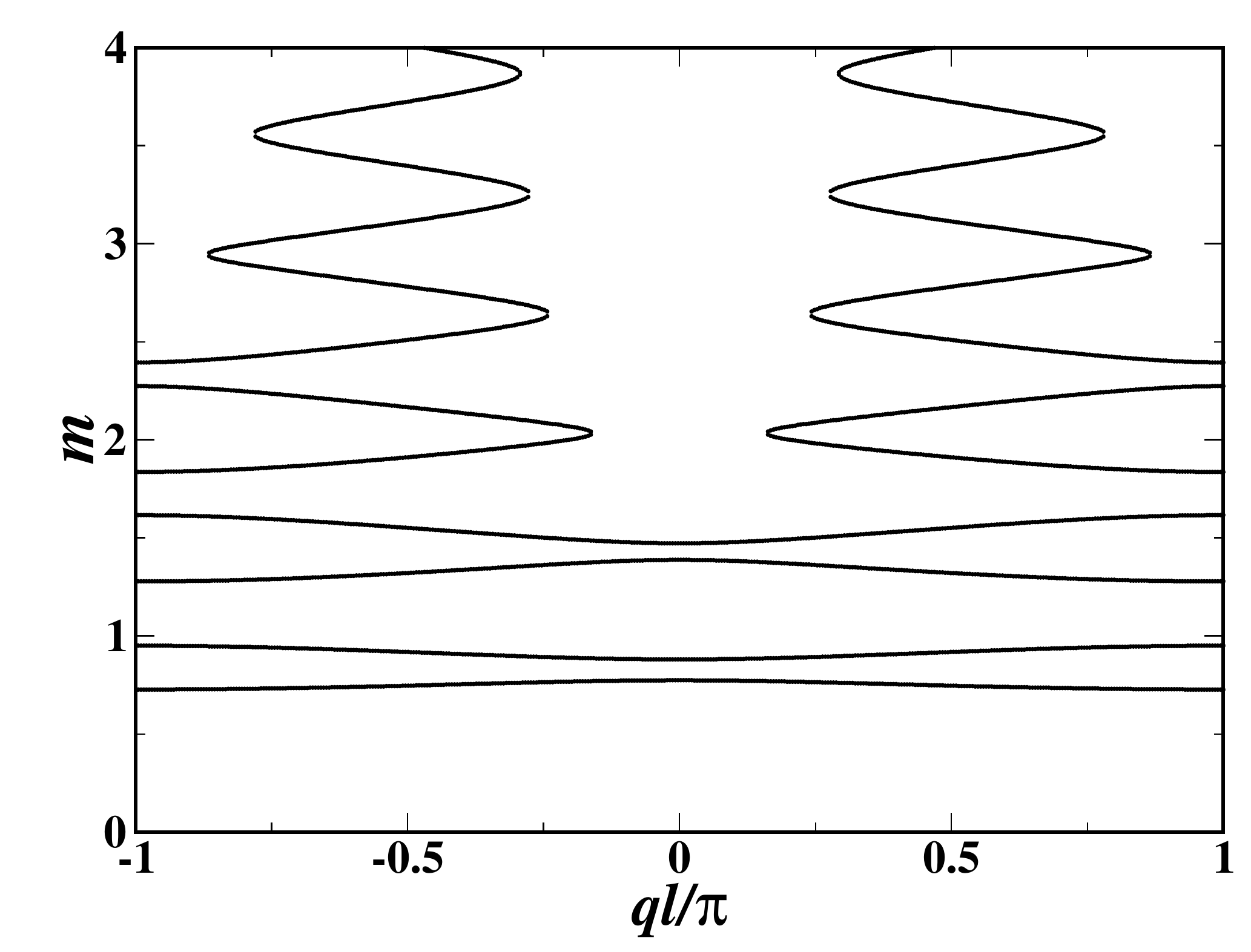,scale=0.55}}
\caption{The lowest mass modes for the case without dilaton and the scalar field $\nu=2$}
\label{v=2}
\end{figure} 
From the above expression we get the condition $\lambda>(8q-4p+3)/4\sqrt{3M^{3}}$
for localization. We must note that all the previous conditions are enclosed
in this one. This expression is the same obtained for the localization
in the case of one brane. To consider the massive modes we must consider the metric defined by Eq. (\ref{metricadilaton}), and the equations of motion for this case is given by
\begin{equation}
\partial_{M}(\sqrt{-g}g^{MN_{1}}\cdots g^{M_{q}N_{q+1}}Y_{N_{1}...N_{q+1}})=0.
\end{equation}
Using the gauge freedom to fix $X_{\mu_{1}\cdots\mu_{q-1}y}=\partial^{\nu}X_{\nu...\mu_{q}}=0$
and the transformations $\frac{dz}{dy}=e^{-3A/4}$ and $\frac{dz}{dy}=e^{-A}$
we arrive to the following expressions for the parameters $b_i$ and $c_i$;
 
\begin{eqnarray}
b_{1}&=&\frac{3}{4},\quad c_{1}=-\left(\frac{\alpha}{2}+\frac{3}{8}\right),\nonumber
\\
b_{2}&=&1,\quad c_{2}=\left(\frac{p}{2}-q\right),
\end{eqnarray}
with $\alpha=(8q-4p-3)/4-\lambda\sqrt{3M^{3}}$. From these the solutions are given by
\begin{eqnarray}
\nu_{1}^{2}&=&\left(\frac{2\alpha}{3}\right)^{2},\nonumber
\\
\nu_{2}^{2}&=&\left(\frac{1+p}{2}-q\right)^{2}.
\end{eqnarray}

\section{Results and discussions}
\begin{figure}
\centerline{\epsfig{figure=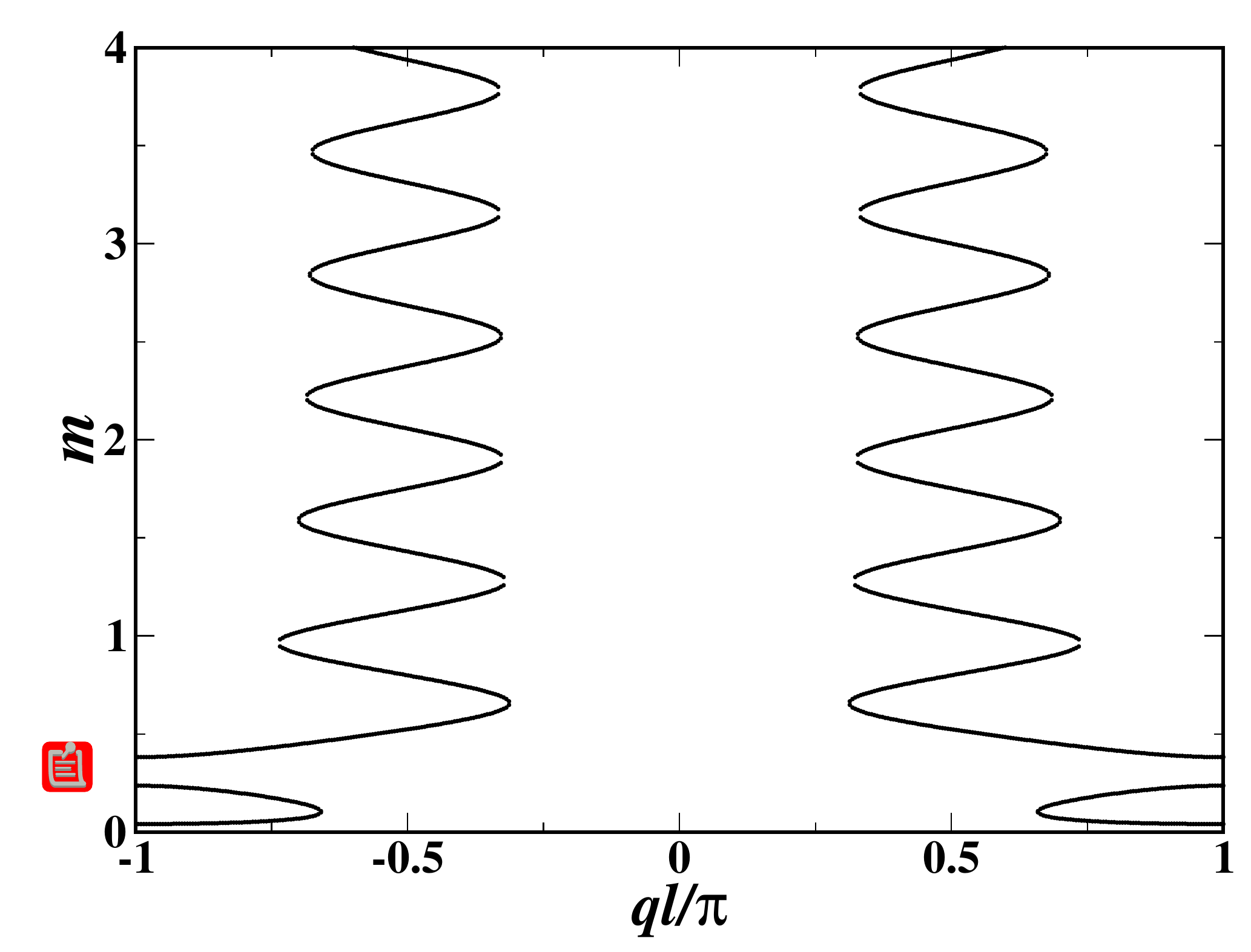,scale=0.55}}
\caption{The mass modes for both the gauge ($\nu=1.0$) and Kalb-Ramond ($\nu=0.0$) fields}\label{v=1}
\end{figure}

The important information one can get from the calculations in previous sections is the allowed values of mass for each field. That can be obtained by solving numerically the equation (\ref{massband}) for the cases with and without dilaton.  For the case without dilaton, we solve Eq. (\ref{massband}) for the scalar field using $\nu=2$, for the gauge field $\nu=1$, and the Kalb-Ramond feld $\nu=0$. The allowed masses for the scalar field are shown in Fig. \ref{v=2} where the lowest mass modes are plotted. As one can see, the lowest mode or the gap of mass has its maximum value  at $q=0$, and for small values of $|q|$ there are only four allowed mass values. The interesting feature about the dispersion relation equation is that due to the property of the Bessel function $j_{-1}=-j_1$ and $n_{-1}=-n_{1}$, the gauge field and the Kalb-Ramond field have the same mass dispersion as shown in Fig. \ref{v=1}. The main characteristic of these fields is that around $q=0$ no mass is allowed. In fact for $m>0.5$ the 
existence of mass particles is restricted to values of $q\approx 0.5\pi/l$. 

\begin{figure}
\centerline{\epsfig{figure=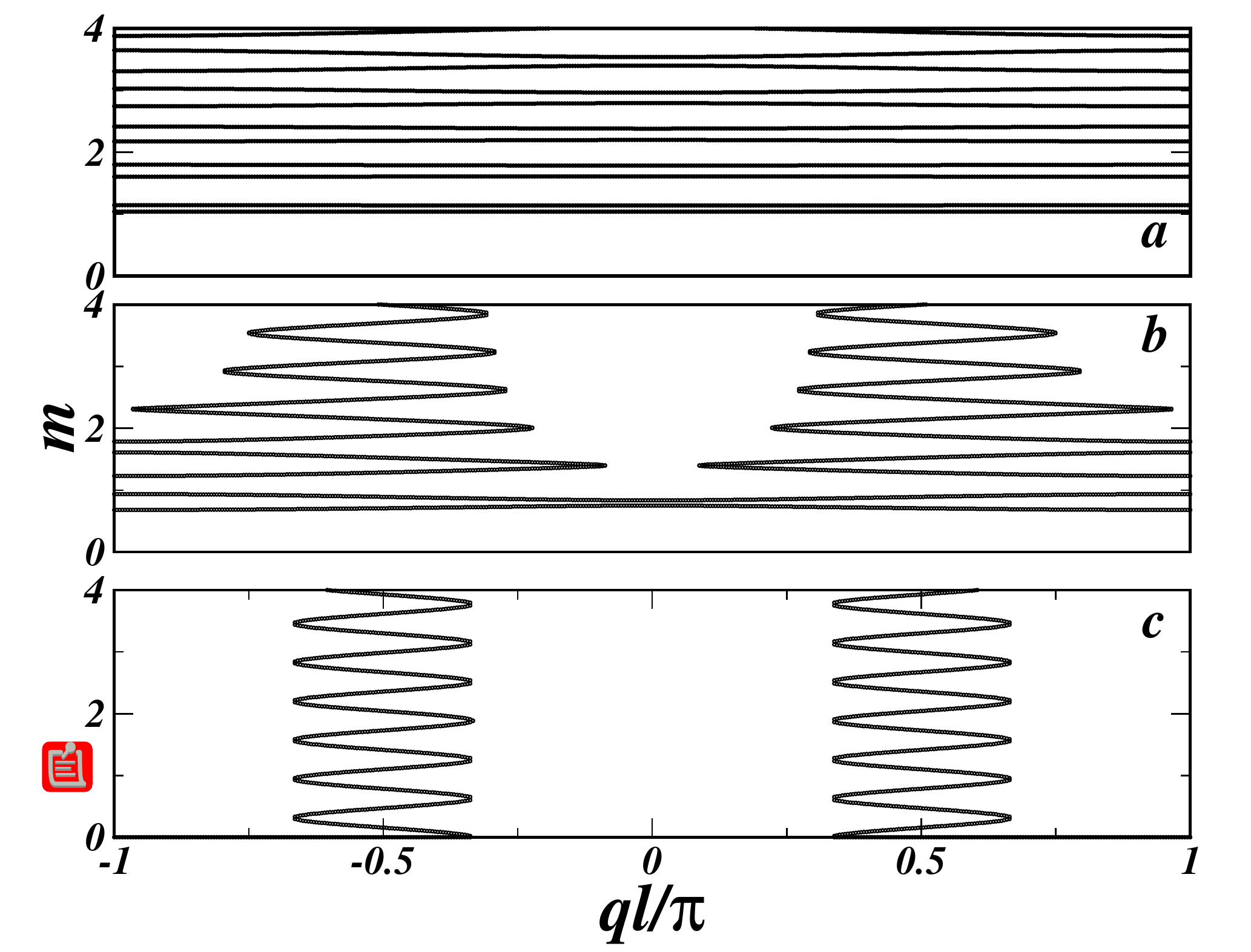,scale=0.55}}
\caption{The mass spectrum for  fields with a dilaton coupling $\lambda=1/\sqrt{(3M^3)}$: a) for the scalar field ($\nu=3.17$), b) for the gauge field ($\nu=1.84$) and c) the Kalb-Ramond field ($\nu=0.5$)}\label{dilaton}
\end{figure}

\begin{figure}[t]
\centerline{\epsfig{figure=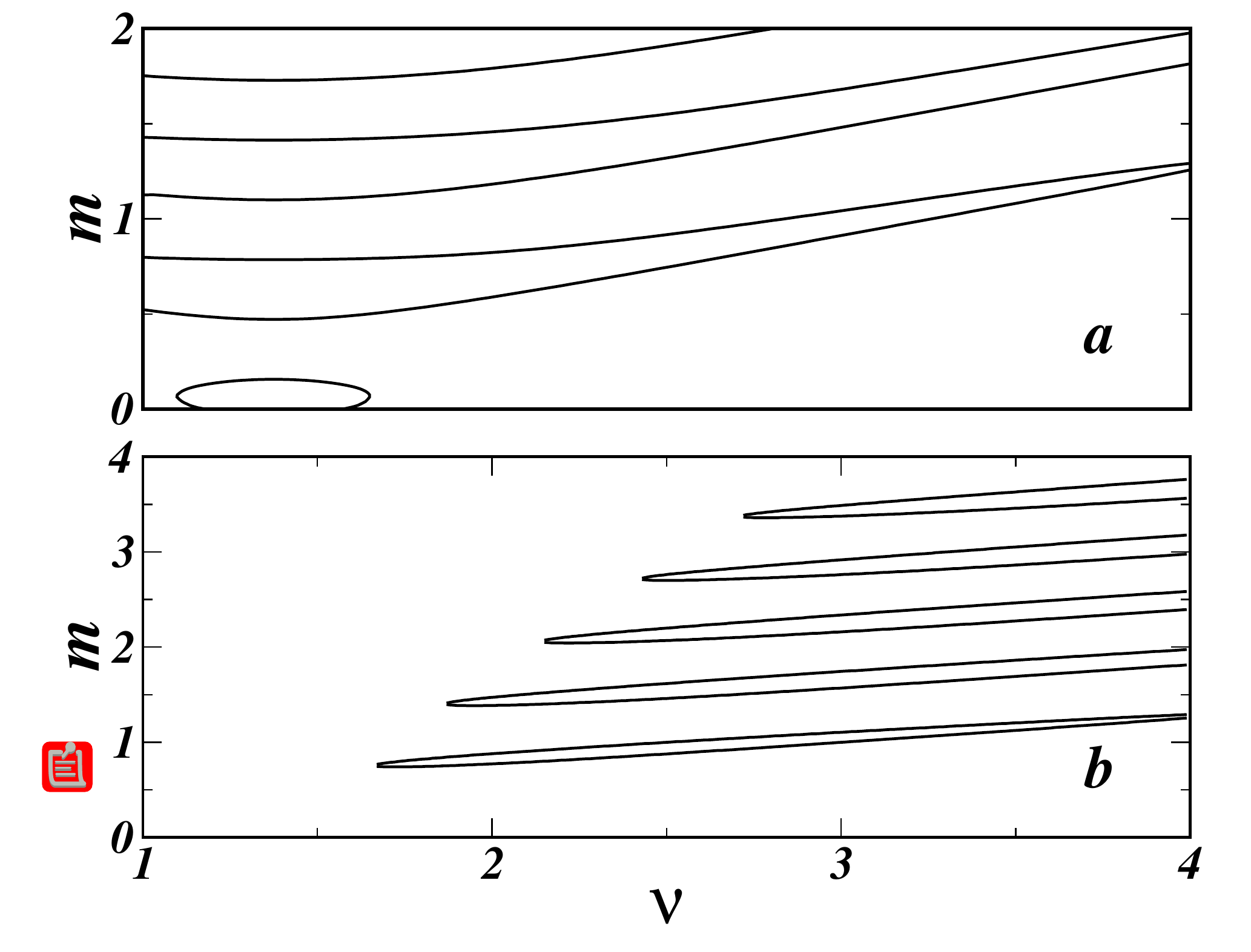,scale=0.55}}
\caption{The mass dispersion against the order of the Bessel function for: a) $q=0.5\pi/l$, and b) $q=0.0$}\label{nu}
\end{figure}

When considering the dilaton, depending on the coupling the  order of the Bessel functions in the dispersion relation can be non-integer. For $\lambda=1/\sqrt{(3M^3)}$ Figure \ref{dilaton} shows the mass dispersion for the $q=0$, $1$, and $2$ forms in panels $a$, $b$, and $c$ respectively.  For the scalar field Fig. \ref{dilaton}$a$ shows a different behavior from the Fig. \ref{v=2} the main effect is now the existence of more mass modes around $q=0$. A similar effect happens for the gauge field, whereas in Fig. \ref{v=2} there was no allowed mass for $q=0$, one can see two possible values of mass in Fig. \ref{dilaton}$b$. The dilaton coupling breaks the degeneracy of the mass modes for the gauge and Kalb-Ramond fields. Now, the two spectra are very different (see Fig. \ref{dilaton}$c$). The main effect of the coupling is to change the order of the Bessel functions. The behavior of the lowest mass modes against the Bessel function order is shown in Fig. \ref{nu}. 

In the Figure \ref{nu}$a$ it is possible to distinguish two regimes: for $\nu>1$ the mass values increases, while for $\nu\le1$ the masses have a parabolic behavior. For $q=0.0$ Fig. \ref{nu}$b$ shows that below $\nu\approx 5/3$ no masses are allowed. These results leads to the fact that controlling the dilaton coupling one can generate or suppress mass modes. A similar effect can be obtained by playing around with the dimension $D$ and the form of the field $q$. For example, in order to get $\nu=5/3$ one can have a dilaton coupling $\lambda=3(3M^3)^{-1/2}/4$ and the condition $q=(2+D)/2$, i.e., only an even dimension is allowed to give an integer $q$ form: for $D=2$ we must have q=2. When considering $\lambda=7(3M^3)^{-1/2}/4$ the condition is $q=(3+D)/2$, in this case only an odd dimension can give an integer $q$. In this case for $D=1$ we can have $q=2$.

\begin{figure}[b]
\centerline{\epsfig{figure=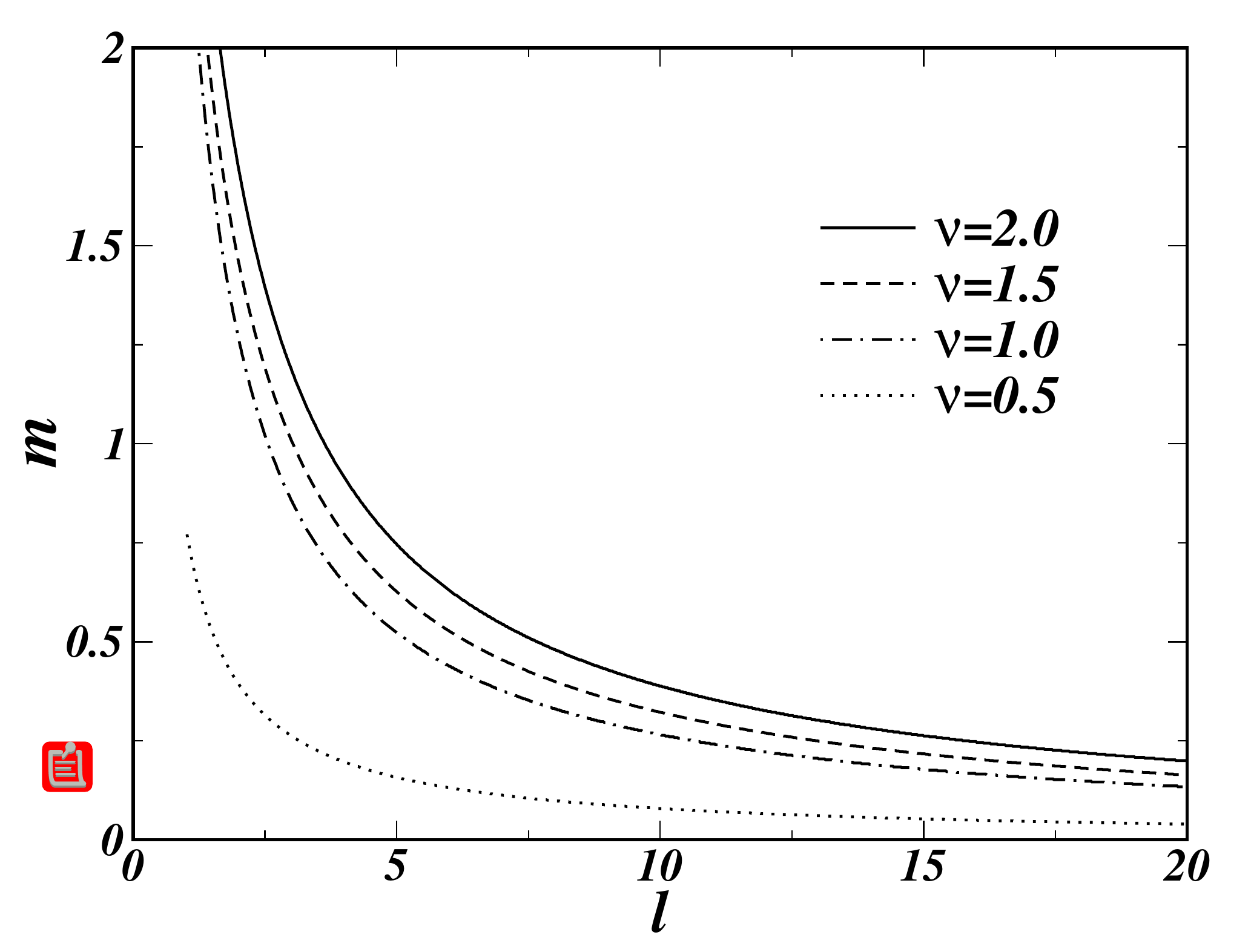,scale=0.55}}
\caption{The lowest mass mode for different values of the Bessel function order $\nu$}\label{length}
\end{figure}

Another interesting aspect of the mass dispersion relation is the lowest mass behavior against the spacing between the branes as shown in Figure \ref{length}. In there, for several values of $\nu$ the mass mode decreases as $m(l)\propto1/l$. For large separations the are not interfering with each other and the system behaves like a single brane and no gap appears for any  any $q$-form or dilaton coupling.

\section{Conclusions}

In this paper we studied aspects of bosonic fields subject to a gravitational background generated by a brane crystal system, where the simple case of branes distributed only along one extra spatial dimension was considered. The behavior of the scalar field, the gauge vector field, the Kalb-Ramond field and general $q$-forms was analyzed. The dispersion relation for the mass was presented for each field mentioned above. Without the dilaton we noted that the Gauge and the Kalb-Ramond fields have the same mass spectrum when $D=5$. However for a lower dimension, for example $D=3$, the scalar field and the gauge field will have the same spectrum. This is an interesting result, since different fields can have the same allowed values of mass for bosonic particles. Even with the introduction of the dilaton coupling, different forms with different couplings can have the same allowed values of mass. Therefore, the main conclusion is the fact different bosonic fields can produce particles with the same mass. Another 
important aspect, is the analogy with the problem of an electron traveling through a periodic potential. In that case, one can calculate the allowed values of the electron's energy where the fields used here make the role of constraints in the potential of the 1D-crystal. That opens the possibility to relate physical effects in condensed matter physics to the physics of extra dimensions.

A natural extension of this work is to study fermions subject to the same background. Besides, it is very important to avoid physical space-time singularities due to the delta function like branes. In this case, we could propose smooth configurations mimicking the crystal. Such configurations can be obtained as approximated multi-kink solutions of a specific scalar field configuration. These are discussions left for future researche.
\section*{Acknowledgments}

We acknowledge the financial support provided by Funda\c c\~ao Cearense de Apoio ao Desenvolvimento Cient\'\i fico e Tecnol\'ogico (FUNCAP), the Conselho Nacional de
Desenvolvimento Cient\'\i fico e Tecnol\'ogico (CNPq) and FUNCAP/CNPq/PRONEX.

This paper is dedicated to the memory of my wife  Isabel Mara (R. R. Landim).


\begin{thebibliography}{99}

\bibitem{Bazeia:2007nd} 
  D.~Bazeia, A.~R.~Gomes and L.~Losano,
  Int.\ J.\ Mod.\ Phys.\ A {\bf 24}, 1135 (2009)
  [arXiv:0708.3530 [hep-th]].

\bibitem{Fonseca:2012bw} 
  R.~C.~Fonseca, F.~A.~Brito and L.~Losano,
  arXiv:1211.0531 [hep-th].

\bibitem{Chumbes:2011zt} 
  A.~E.~R.~Chumbes, J.~M.~Hoff da Silva and M.~B.~Hott,
  Phys.\ Rev.\ D {\bf 85}, 085003 (2012)
  [arXiv:1108.3821 [hep-th]].
  
\bibitem{Cunha:2011yk} 
  M.~S.~Cunha and H.~R.~Christiansen,
  Phys.\ Rev.\ D {\bf 84}, 085002 (2011)
  [arXiv:1109.3486 [hep-th]].

\bibitem{Almeida:2009jc} 
  C.~A.~S.~Almeida, M.~M.~Ferreira, Jr., A.~R.~Gomes and R.~Casana,
  Phys.\ Rev.\ D {\bf 79}, 125022 (2009)
  [arXiv:0901.3543 [hep-th]].

  
\bibitem{Liu:2009ve} 
  Y.~-X.~Liu, J.~Yang, Z.~-H.~Zhao, C.~-EFu and Y.~-S.~Duan,
  Phys.\ Rev.\ D {\bf 80}, 065019 (2009)
  [arXiv:0904.1785 [hep-th]].
  
\bibitem{Liu:2009mga} 
  Y.~-X.~Liu, H.~-T.~Li, Z.~-H.~Zhao, J.~-X.~Li and J.~-R.~Ren,
  JHEP {\bf 0910}, 091 (2009)
  [arXiv:0909.2312 [hep-th]].
  
\bibitem{Cruz:2009ne} 
  W.~T.~Cruz, M.~O.~Tahim and C.~A.~S.~Almeida,
  Europhys.\ Lett.\  {\bf 88}, 41001 (2009)
  [arXiv:0912.1029 [hep-th]].
  
\bibitem{Zhao:2010mk} 
  Z.~-H.~Zhao, Y.~-X.~Liu, H.~-T.~Li and Y.~-Q.~Wang,
  Phys.\ Rev.\ D {\bf 82}, 084030 (2010)
  [arXiv:1004.2181 [hep-th]].
  
\bibitem{Li:2010dy} 
  H.~-T.~Li, Y.~-X.~Liu, Z.~-H.~Zhao and H.~Guo,
  Phys.\ Rev.\ D {\bf 83}, 045006 (2011)
  [arXiv:1006.4240 [hep-th]].
 
  
\bibitem{Cruz:2011kj} 
  W.~T.~Cruz, A.~R.~Gomes and C.~A.~S.~Almeida,
  Europhys.\ Lett.\  {\bf 96}, 31001 (2011)
  [arXiv:1110.3104 [hep-th]].
  
  
\bibitem{Alencar:2012en} 
  G.~Alencar, R.~R.~Landim, M.~O.~Tahim and R.~N.~C.~Filho,
  arXiv:1207.3054 [hep-th].
  
\bibitem{Landim:2011ts} 
  R.~R.~Landim, G.~Alencar, M.~O.~Tahim and R.~N.~Costa Filho,
  JHEP {\bf 1202}, 073 (2012)
  [arXiv:1110.5855 [hep-th]].
  
\bibitem{Landim:2011ki} 
  R.~R.~Landim, G.~Alencar, M.~O.~Tahim and R.~N.~Costa Filho,
  JHEP {\bf 1108}, 071 (2011)
  [arXiv:1105.5573 [hep-th]].
  
\bibitem{Csaki:2002gy} 
  C.~Csaki, J.~Erlich and J.~Terning,
  Phys.\ Rev.\ D {\bf 66}, 064021 (2002)
  [hep-ph/0203034].

\bibitem{Csaki:1999mp} 
  C.~Csaki, M.~Graesser, L.~Randall and J.~Terning,
  Phys.\ Rev.\ D {\bf 62}, 045015 (2000)
  [hep-ph/9911406].

\bibitem{Csaki:1999jh} 
  C.~Csaki, M.~Graesser, C.~F.~Kolda and J.~Terning,
  Phys.\ Lett.\ B {\bf 462}, 34 (1999)
  [hep-ph/9906513].
  
  

\bibitem{Kaloper:1999et} 
  N.~Kaloper,
  Phys.\ Lett.\ B {\bf 474}, 269 (2000)
  [hep-th/9912125].
  
\bibitem{Kaloper:2004cy} 
  N.~Kaloper,
  JHEP {\bf 0405}, 061 (2004)
  [hep-th/0403208].
  
\bibitem{Oda:1999di} 
  I.~Oda,
  Phys.\ Lett.\ B {\bf 480}, 305 (2000)
  [hep-th/9908104].
 
\bibitem{Aldazabal:2000cn} 
  G.~Aldazabal, S.~Franco, L.~E.~Ibanez, R.~Rabadan and A.~M.~Uranga,
  JHEP {\bf 0102}, 047 (2001)
  [hep-ph/0011132].
  
\bibitem{Ibanez:2001nd}
  L.~E.~Ibanez, F.~Marchesano and R.~Rabadan,
  JHEP {\bf 0111} (2001) 002
  [hep-th/0105155].
  
\bibitem{Blumenhagen:2001te} 
  R.~Blumenhagen, B.~Kors, D.~Lust and T.~Ott,
  Nucl.\ Phys.\ B {\bf 616}, 3 (2001)
  [hep-th/0107138].
  
\bibitem{Blumenhagen:2005mu} 
  R.~Blumenhagen, M.~Cvetic, P.~Langacker and G.~Shiu,
  Ann.\ Rev.\ Nucl.\ Part.\ Sci.\  {\bf 55}, 71 (2005)
  [hep-th/0502005].
  
\bibitem{Lust:2004ks} 
  D.~Lust,
  Class.\ Quant.\ Grav.\  {\bf 21}, S1399 (2004)
  [hep-th/0401156].
  
  

  

\bibitem{Polchinski:1998rq}
  J.~Polchinski,
``String theory. Vol. 1: An introduction to the bosonic string,''
{SPIRES
entry} {\it  Cambridge, UK: Univ. Pr. (1998) 402 p}

\bibitem{Polchinski:1998rr}
  J.~Polchinski,
``String theory. Vol. 2: Superstring theory and beyond,''
{SPIRES entry} {\it  Cambridge, UK: Univ. Pr. (1998) 531 p}

\bibitem{VanNieuwenhuizen:1981ae}
  P.~Van Nieuwenhuizen,
  Phys.\ Rept.\  {\bf 68}, 189 (1981).

\bibitem{Smailagic:1999qw}
  A.~Smailagic and E.~Spallucci,
  Phys.\ Rev.\  D {\bf 61}, 067701 (2000)
  [arXiv:hep-th/9911089].

\bibitem{Smailagic:2000hr}
  A.~Smailagic and E.~Spallucci,
  Phys.\ Lett.\  B {\bf 489}, 435 (2000)
  [arXiv:hep-th/0008094].

\bibitem{Oda:1989tq}
  I.~Oda and S.~Yahikozawa,
  ``Linking Numbers And Variational Method,''
  Phys.\ Lett.\  B {\bf 238}, 272 (1990).
\bibitem{Landim:2000ir} 
  R.~R.~Landim and C.~A.~S.~Almeida,
  Phys.\ Lett.\ B {\bf 504}, 147 (2001)
  [hep-th/0010050].
\bibitem{Landim:2001zu} 
  R.~R.~Landim,
  Phys.\ Lett.\ B {\bf 542}, 160 (2002)
  [hep-th/0108242].
  
\bibitem{Nakahara:2003nw}
  M.~Nakahara,
  ``Geometry, topology and physics,''
{\it  Boca Raton, USA: Taylor} \& {\it Francis (2003) 573 p}

\bibitem{Kachru:2002he}
  S.~Kachru, M.~B.~Schulz and S.~Trivedi,
  ``Moduli stabilization from fluxes in a simple IIB orientifold,''
  JHEP {\bf 0310}, 007 (2003)
  [arXiv:hep-th/0201028].

\bibitem{Antoniadis:2004pp}
  I.~Antoniadis and T.~Maillard,
  ``Moduli stabilization from magnetic fluxes in type I string theory,''
  Nucl.\ Phys.\  B {\bf 716}, 3 (2005)
  [arXiv:hep-th/0412008].

\bibitem{Balasubramanian:2005zx}
  V.~Balasubramanian, P.~Berglund, J.~P.~Conlon and F.~Quevedo,
  ``Systematics of Moduli Stabilisation in Calabi-Yau Flux Compactifications,''
  JHEP {\bf 0503}, 007 (2005)
  [arXiv:hep-th/0502058].
  
 
\bibitem{Alencar:2010vk} 
  G.~Alencar, R.~R.~Landim, M.~O.~Tahim, K.~C.~Mendes, R.~R.~Landim, M.~O.~Tahim, R.~N.~C.~Filho and K.~C.~Mendes,
  Europhys.\ Lett.\  {\bf 93}, 10003 (2011)
  [arXiv:1009.1183 [hep-th]].

 
\bibitem{Mukhopadhyaya:2004cc}
  B.~Mukhopadhyaya, S.~Sen, S.~Sen and S.~SenGupta,
  ``Bulk Kalb-Ramond field in Randall Sundrum scenario,''
  Phys.\ Rev.\  D {\bf 70}, 066009 (2004)
  [arXiv:hep-th/0403098]

\bibitem{Mukhopadhyaya:2007jn}
  B.~Mukhopadhyaya, S.~Sen and S.~SenGupta,
  ``Bulk antisymmetric tensor fields in a Randall-Sundrum model,''
  Phys.\ Rev.\  D {\bf 76}, 121501 (2007)
  [arXiv:0709.3428 [hep-th]].


\bibitem{DeRisi:2007dn}
  G.~De Risi,
  ``Bouncing cosmology from Kalb-Ramond Braneworld,''
  Phys.\ Rev.\  D {\bf 77}, 044030 (2008)
  [arXiv:0711.3781 [hep-th]].

\bibitem{Mukhopadhyaya:2009gp}
  B.~Mukhopadhyaya, S.~Sen and S.~SenGupta,
  ``A Randall-Sundrum scenario with bulk dilaton and torsion,''
  Phys.\ Rev.\  D {\bf 79}, 124029 (2009)
  [arXiv:0903.0722 [hep-th]].


\bibitem{Alencar:2010mi} 
  G.~Alencar, M.~O.~Tahim, R.~R.~Landim, C.~R.~Muniz and R.~N.~Costa Filho,
  Phys.\ Rev.\ D {\bf 82}, 104053 (2010)
  [arXiv:1005.1691 [hep-th]].



\bibitem{Kehagias:2000au}
  A.~Kehagias and K.~Tamvakis,
  Phys.\ Lett.\ B {\bf 504} (2001) 38
  [hep-th/0010112].


\bibitem{Alencar:2010hs} 
  G.~Alencar, R.~R.~Landim, M.~O.~Tahim, C.~R.~Muniz and R.~N.~Costa Filho,
  Phys.\ Lett.\ B {\bf 693}, 503 (2010)
  [arXiv:1008.0678 [hep-th]].
  
\bibitem{Landim:2010pq} 
  R.~R.~Landim, G.~Alencar, M.~O.~Tahim, M.~A.~M.~Gomes and R.~N.~Costa Filho,
  Europhys.\ Lett.\  {\bf 97}, 20003 (2012)
  [arXiv:1010.1548 [hep-th]].

\bibitem{Fu:2012sa} 
  C.~-EFu, Y.~-X.~Liu, K.~Yang and S.~-W.~Wei,
  JHEP {\bf 1210}, 060 (2012)
  [arXiv:1207.3152 [hep-th]].
  
  
\bibitem{ArkaniHamed:1999zg} 
  N.~Arkani-Hamed, S.~Dimopoulos, G.~R.~Dvali and N.~Kaloper,
  JHEP {\bf 0012}, 010 (2000)
  [hep-ph/9911386].
  
\end{thebibliography}
\end{document}